# A new class of nonreciprocal spin waves on the edges of 2D antiferromagnetic honeycomb nanoribbons

D. Ghader*[(1)] and A. Khater[(2,3)]

[1] College of Engineering and Technology, American University of the Middle East, Eqaila, Kuwait

[2] Department of Theoretical Physics, Institute of Physics, Jan Dlugosz University, Am. Armii Krajowej 13/15, Czestochowa, Poland

[3] Department of Physics, University du Maine, 72085 Le Mans, France

**Antiferromagnetic two-dimensional (2D) materials are currently under intensive theoretical and experimental investigations in view of their potential applications in antiferromagnet-based magnonic and spintronic devices. Recent experimental studies revealed the importance of magnetic anisotropy and of Dzyaloshinskii-Moriya interactions (DMI) on the ordered ground state and the magnetic excitations in these materials. In this work we present a robust classical field theory approach to study the effects of anisotropy and the DMI on the edge and bulk spin waves in 2D antiferromagnetic nanoribbons. We predict the existence of a new class of nonreciprocal edge spin waves, characterized by opposite polarizations in counter-propagation. These novel edge spin waves are induced by the DMI and are fundamentally different from conventional nonreciprocal spin waves for which the polarization is independent of the propagation direction. We further analyze the effects of the edge structures on the magnetic excitations for these systems. In particular, we show that anisotropic bearded edge nanoribbons act as topologically trivial magnetic insulators with potentially interesting applications in magnonics. Our results constitute an important finding for current efforts seeking to establish unconventional magnonic devices utilizing spin wave polarization.**

The polarization of a spin wave is determined by the precessing direction of the magnetization and constitutes an important additional intrinsic degree of freedom, beside the spin wave amplitude and phase. Spin wave theory in a collinear antiferromagnet, composed of two magnetic lattices with opposite magnetization, is known to yield two bulk modes. The precession frequencies of these modes are characterized by opposite contributions from the exchange interaction [1-3]. As a consequence, the sublattice magnetizations precess clockwise in one of the modes and anti-clockwise in the other [3]. The spin wave modes in an antiferromagnet are hence characterized by opposite polarizations, conventionally termed as right-handed and left-handed spin waves.

The important advantages for utilizing spin wave polarization to encode and process information in antiferromagnet-based magnonics has recently received significant attention [4-8]. The Dzyaloshinskii-Moriya interactions (DMI) [9, 10] was highlighted as a key ingredient to realize polarization-based magnonic devices. In particular, antiferromagnetic domain walls with DMI have been proposed as spin wave polarizer, retarder and transistor [4, 5].

Unlike the symmetric Heisenberg exchange interaction, the DMI is antisymmetric and breaks the reflection symmetry in a magnetic material. Consequently, the DMI contribution to the spin waves dispersion curves can be asymmetric, giving rise to nonreciprocal spin wave excitations. Nonreciprocal spin waves are characterized by different frequencies and amplitudes when propagating in opposite directions. These may occur owing to several factors including DMI, inhomogeneous magnetization, externally applied magnetic fields, and asymmetric magnetic anisotropies on the surfaces of a magnetic film. The polarization of conventional nonreciprocal spin waves, however, is independent of the propagation direction. Nonreciprocal bulk, surface, interface and domain wall spin waves in magnetic films and layered structures have been widely explored for fundamental and technological interests [11-25].

Recently, a new research field emerged exploring magnetic excitations in two-dimensional (2D) and quasi-2D-layered van der Waals materials [26-52]. Given their novel characteristics, these recently discovered materials are expected to open new routes in magnonics and spintronics. Experimental studies on quasi-2D ferromagnets and antiferromagnets revealed the importance of magnetic anisotropy and DMI on spin excitations in these materials [36, 44, 47, 50]. Several interesting approaches have also been proposed to tune the magnetic anisotropy and the DMI [49, 53-57].

Bounded 2D magnetic materials present edge spin waves, considered as the 1D analogue of surface spin waves in magnetic thin films. The interest in surface spin waves is naturally extended to edge spin waves [26, 39, 40, 48, 51, 52], as fundamental phenomena with potentials for pioneering applications. Despite the intensive research on magnetic excitations in 2D materials, the effects of the DMI and magnetic anisotropy on edge spin waves in bounded 2D antiferromagnets have not yet received the deserved attention and remain open to be discovered.

In the present work we develop a classical field approach, with appropriate boundary conditions, to explore the combined effect of DMI and magnetic anisotropy on the edge and bulk spin waves in 2D antiferromagnetic honeycomb nanoribbons. We predict a new class of nonreciprocal edge spin waves characterized by opposite (right and left-handed) polarizations when propagating in opposite directions. The study further reveals unconventional and interesting features for these spin waves, promoting 2D honeycomb antiferromagnets as potentially promising for applications in magnonics. In particular, we find that anisotropic nanoribbons with bearded edge boundaries host low-energy edge spin waves that can be excited separately from the bulk modes, hence acting as magnetic insulators. In the presence of the DMI, the low-energy edge spin waves propagate exclusively on one edge of the nanoribbon, with controllable propagation direction and polarization. Nanoribbons with bearded edges are also characterized by DMI induced unidirectional high-energy edge spin waves, above the propagation band, with remarkably large group velocities near the Brillouin Zone (BZ) boundary.

## Results

**Bulk spin dynamics.** Representative bulk sites in the 2D antiferromagnetic honeycomb nanoribbon are presented schematically in Fig.1. In the Néel antiferromagnetic ordering state, the spins on A (blue) and B (red) honeycomb sublattices are conventionally assumed to be aligned parallel and antiparallel to the z-axis. The interaction in the 2D antiferromagnetic honeycomb nanoribbon is described by a semi-classical Heisenberg Hamiltonian expressed as follows

$$\mathcal{H} = J \sum_{\langle \vec{r},\vec{\delta} \rangle} \left[ \vec{S}_{\parallel}(\vec{r},t) \cdot \vec{S}_{\parallel}(\vec{r}+\vec{\delta},t) + \gamma S_z(\vec{r},t) S_z(\vec{r}+\vec{\delta},t) \right]$$
$$+ \sum_{\langle \vec{r},\vec{\delta}' \rangle} \vec{D}(\vec{r},\vec{r}+\vec{\delta}') \cdot \left[ \vec{S}(\vec{r},t) \times \vec{S}(\vec{r}+\vec{\delta}',t) \right]$$

Here, $\vec{r} = x\hat{x} + y\hat{y}$ is the position vector of a site on the honeycomb lattice, $t$ is time, $\vec{\delta}$ and $\vec{\delta}'$ are the position vectors of a nearest neighbor and a next nearest neighbor respectively. The vector $\vec{S}_{\parallel} = S_x \hat{x} + S_y \hat{y}$ represents the spin component in the plane of the honeycomb lattice. $J$ is the exchange constant and $\gamma \geq 1$ is the anisotropy parameter along the z-axis.

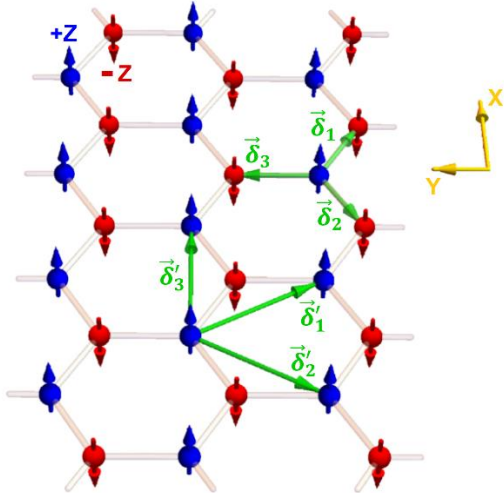

**Fig. 1:** Schematic representation of bulk sites in a 2D antiferromagnetic honeycomb nanoribbon. The nanoribbon is finite along the y-axis and infinite along the x-axis. The green vectors connect a bulk A-site to its nearest and next nearest neighbors.

The first and second terms in the Hamiltonian respectively represent the exchange and DM interactions. The DMI vector is given by $\vec{D}(\vec{r},\vec{r}+\vec{\delta}') = D_z(\vec{r},\vec{r}+\vec{\delta}')\hat{z}$ with $D_z(\vec{r},\vec{r}+\vec{\delta}') = \pm D$. The parameter $D$ determines the strength of the DMI, whereas the orientation of $\vec{D}$ in the honeycomb lattice is determined in the conventional way from the local geometry for a nonzero output with reference to the ribbon spin vectors [40, 48, 58].

For the exchange part, the nearest neighbors summation is over $\vec{\delta}_1 = \frac{a}{2}\hat{x} - \frac{a}{2\sqrt{3}}\hat{y}$, $\vec{\delta}_2 = -\frac{a}{2}\hat{x} - \frac{a}{2\sqrt{3}}\hat{y}$ and $\vec{\delta}_3 = \frac{a}{\sqrt{3}}\hat{y}$ for a bulk A-site, and over $-\vec{\delta}_1$, $-\vec{\delta}_2$ and $-\vec{\delta}_3$ for a bulk B-site. The next nearest neighbors summation for the DMI on both A and B bulk sites runs over $\pm\vec{\delta}_1'$, $\pm\vec{\delta}_2'$ and $\pm\vec{\delta}_3'$. Here, $\vec{\delta}_1' = \frac{a}{2}\hat{x} - \frac{\sqrt{3}\,a}{2}\hat{y}$, $\vec{\delta}_2' = -\frac{a}{2}\hat{x} - \frac{\sqrt{3}\,a}{2}\hat{y}$ and $\vec{\delta}_3' = a\hat{x}$. The vectors are illustrated in Fig. 1.

We start with a trick to unify the treatment of the exchange and DM interactions within the classical field formalism. The DMI between an A-site and any of its next nearest neighbors can be rewritten as

$$\vec{D}^A(\vec{r},\vec{r}+\vec{\delta}') \cdot [\vec{S}^A(\vec{r},t) \times \vec{S}^A(\vec{r}+\vec{\delta}',t)] = D_z^A(\vec{r},\vec{r}+\vec{\delta}')\,\vec{S}^A(\vec{r},t) \cdot \vec{S}_D^A(\vec{r}+\vec{\delta}',t) \qquad (1)$$

with $\vec{S}_D^A(\vec{r}+\vec{\delta}',t) = S_y^A(\vec{r}+\vec{\delta}',t)\,\hat{x} - S_x^A(\vec{r}+\vec{\delta}',t)\,\hat{y}$

A similar equation holds for a B-site. With equation 1, one can use the standard classical field theory formalism [1-3, 11, 51, 52, 59-63] to determine the effective fields $\vec{H}^A$ and $\vec{H}^B$ acting on the sublattice magnetizations $\vec{M}^A$ and $\vec{M}^B$ respectively. These fields can now be expressed as

$$\vec{H}^{A/B} = \vec{H}_{ex}^{A/B} + \vec{H}_{DM}^{A/B} \qquad (2)$$

with exchange fields

$$\vec{H}_{ex}^A = J\sum_{\vec{\delta}} \vec{M}_\parallel^B(\vec{r}+\vec{\delta},t) - 3\gamma J M\,\hat{z} \qquad (3a)$$

$$\vec{H}_{ex}^B = J\sum_{\vec{\delta}} \vec{M}_\parallel^A(\vec{r}+\vec{\delta},t) + 3\gamma J M\,\hat{z} \qquad (3b)$$

and DMI fields

$$\vec{H}_{DM}^A = \sum_{\vec{\delta}'} D_z^A(\vec{r},\vec{r}+\vec{\delta}')\,\vec{M}_D^A(\vec{r}+\vec{\delta}',t) \qquad (3c)$$

$$\vec{H}_{DM}^B = \sum_{\vec{\delta}'} D_z^B(\vec{r},\vec{r}+\vec{\delta}')\,\vec{M}_D^B(\vec{r}+\vec{\delta}',t) \qquad (3d)$$

In equations 3, we have used $M = M_z^A = -M_z^B$, $\vec{M}_\parallel^A = M_x^A \hat{x} + M_y^A \hat{y}$ and $\vec{M}_D^A = M_y^A \hat{x} - M_x^A \hat{y}$. The vectors $\vec{M}_\parallel^B$ and $\vec{M}_D^B$ have similar forms.

The nanoribbon is considered finite along the y-direction; the solutions for the magnetization components are hence assumed of the form [11, 51, 52, 59-61, 64-68]

$$M_x^A = e^{i(\omega t - k_x x)}[A_x^+ e^{qy} + A_x^- e^{-qy}] \quad (4a)$$
$$M_y^A = e^{i(\omega t - k_x x)}[A_y^+ e^{qy} + A_y^- e^{-qy}] \quad (4b)$$
$$M_x^B = e^{i(\omega t - k_x x)}[B_x^+ e^{qy} + B_x^- e^{-qy}] \quad (4c)$$
$$M_y^B = e^{i(\omega t - k_x x)}[B_y^+ e^{qy} + B_y^- e^{-qy}] \quad (4d)$$

In equations 4, $q$ is a phase factor along the finite y-direction. Its real and imaginary values correspond respectively to evanescent (edge) and propagating (bulk) spin waves. $k_x$ is the continuous wave vector along the infinite x-direction of the nanoribbon.

Substituting equations 4 in equations 3 yields the effective fields as

$$\vec{H}^A = \phi\{J(B_x^+ e^{qy} f_{ex}^+ + B_x^- e^{-qy} f_{ex}^-) + 4iDf_{DM}(A_y^+ e^{qy} + A_y^- e^{-qy})\}\hat{x}$$
$$+ \phi\{J(B_y^+ e^{qy} f_{ex}^+ + B_y^- e^{-qy} f_{ex}^-) - 4iDf_{DM}(A_x^+ e^{qy} + A_x^- e^{-qy})\}\hat{y} - 3\gamma JM \hat{z} \quad (5a)$$

$$\vec{H}^B = \phi\{J(A_x^+ e^{qy} f_{ex}^- + A_x^- e^{-qy} f_{ex}^+) - 4iDf_{DM}(B_y^+ e^{qy} + B_y^- e^{-qy})\}\hat{x}$$
$$+ \phi\{J(A_y^+ e^{qy} f_{ex}^- + A_y^- e^{-qy} f_{ex}^+) + 4iDf_{DM}(B_x^+ e^{qy} + B_x^- e^{-qy})\}\hat{y} + 3\gamma JM \hat{z} \quad (5b)$$

with

$\phi = e^{i(\omega t - k_x x)}$, $f_{ex}^+ = e^{\frac{a}{\sqrt{3}}q} + 2e^{-\frac{a}{2\sqrt{3}}q}\cos(\frac{k_x a}{2})$, $f_{ex}^- = e^{-\frac{a}{\sqrt{3}}q} + 2e^{\frac{a}{2\sqrt{3}}q}\cos(\frac{k_x a}{2})$, and $f_{DM} = \sin(\frac{k_x a}{2})[\cos(\frac{k_x a}{2}) - \cosh(\frac{\sqrt{3}qa}{2})]$

In the classical field theory, the magnetization dynamics for the 2 sublattices are described by the Bloch (or Landau-Lifshitz) equations of motion, $\partial_t \vec{M}^{A/B} = \lambda \vec{M}^{A/B} \times \vec{H}^{A/B}$. With the help of equations 4 and 5, the Bloch equations yield

$$\left(-\Omega + \frac{4D}{J}f_{DM} + 3\gamma\right)A^+ + f_{ex}^+ B^+ = 0 \tag{6a}$$

$$\left(-\Omega + \frac{4D}{J}f_{DM} + 3\gamma\right)A^- + f_{ex}^+ B^- = 0 \tag{6b}$$

$$\left(-\Omega + \frac{4D}{J}f_{DM} - 3\gamma\right)B^+ - f_{ex}^- A^+ = 0 \tag{6c}$$

$$\left(-\Omega + \frac{4D}{J}f_{DM} - 3\gamma\right)B^- - f_{ex}^- A^- = 0 \tag{6d}$$

with the normalized frequency $\Omega$ defined as $\Omega = \frac{\omega}{\lambda J M}$, $\lambda$ is the gyromagnetic ratio, $A^\pm = A_x^\pm + iA_y^\pm$, and $B^\pm = B_x^\pm + iB_y^\pm$. Details on the derivation of equations 6 are presented in Supplementary Note 1.

The condition that the linear system formed from 6a and 6c (equivalently 6b and 6d) admits nonzero solutions yields a dispersion equation, quadratic in $\Omega$, as follows

$$\left(\Omega - \frac{4D}{J}f_{DM}\right)^2 - 9\gamma^2 = -f_{ex}^+ f_{ex}^- \tag{7}$$

with two solutions,

$$\Omega_\pm = \frac{4D}{J}f_{DM} \pm \sqrt{9\gamma^2 - f_{ex}^+ f_{ex}^-} \tag{8}$$

corresponding to 2 spin wave modes with opposite polarizations [1-3].

Even in the presence of the DMI, the solutions $\Omega_+$ and $\Omega_-$ for propagating modes are respectively positive and negative throughout the BZ. The solution $\Omega_-$ is hence excluded for propagating modes, since the spin wave excitation energy is required to be positive [69]. Consequently, $\Omega_+$ is generally considered as the unique energy dispersion relation for propagating spin waves in an antiferromagnet [11, 51, 52, 58, 59, 61]. Remarkably, we will prove that this conclusion cannot be generalized to the edge spin waves in 2D antiferromagnetic honeycomb nanoribbons with nonzero DMI. In particular, we show that both solutions $\Omega_+$ and $\Omega_-$ conspire to form the positive energy dispersion curve for these edge spin waves.

**Boundary conditions.** We consider honeycomb nanoribbons with zigzag edge (ZE) and bearded edge (BE) boundaries as in Fig. 2. The edges are at $y = \pm d$.

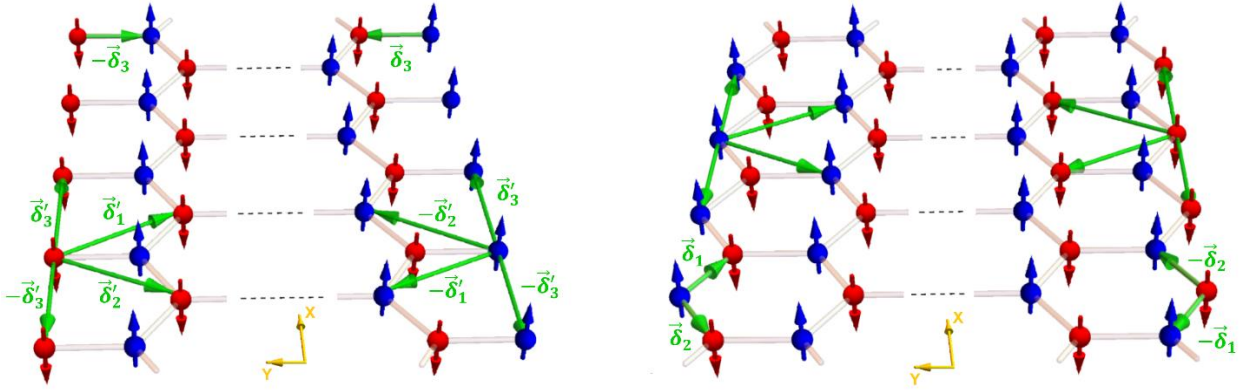

**Fig. 2:** The figure illustrates the reduced number of nearest and next nearest neighbors for edge sites in honeycomb nanoribbons with bearded edge (left) and zigzag edge (right) boundaries.

Boundary conditions for exchange spin waves in magnetic films were first formulated by Rado and Weertman [70]. Later, Stamps and Camley [59] formulated equivalent exchange boundary conditions based on the requirement that the precession frequency of a boundary spin should match that of a bulk spin. Equivalently, boundary spins are required to satisfy the bulk equations of motion [11, 51, 52, 59-61, 64-68] which yields an elegant boundary condition equation

$$\vec{M}_e^{A/B} \times (\vec{H}_b^{A/B} - \vec{H}_e^{A/B}) = \vec{0} \tag{9}$$

In equation 9, the subscripts $e$ and $b$ respectively stand for edge and the bulk. As illustrated in Fig. 2, the number of nearest and next nearest neighbors for an edge site are reduced compared to the bulk site. The edge effective fields hence differ from their bulk counterparts.

Equation 9 developed on the right and left edges of the nanoribbon yields two linear equations in the coefficients $\{A^+, A^-, B^+, B^-\}$, namely

$$e^{qd}\left[\gamma - \frac{2D}{J}\sin\left(\frac{k_x a}{2}\right)e^{\frac{\sqrt{3}a}{2}q}\right]A^+ + e^{-qd}\left[\gamma - \frac{2D}{J}\sin\left(\frac{k_x a}{2}\right)e^{-\frac{\sqrt{3}a}{2}q}\right]A^- + e^{q\left(d+\frac{a}{\sqrt{3}}\right)}B^+ + e^{-q\left(d+\frac{a}{\sqrt{3}}\right)}B^- = 0 \tag{10a}$$

$$e^{-q\left(d+\frac{a}{\sqrt{3}}\right)}A^+ + e^{q\left(d+\frac{a}{\sqrt{3}}\right)}A^- + e^{-qd}\left[\gamma + \frac{2D}{J}\sin\left(\frac{k_x a}{2}\right)e^{-\frac{\sqrt{3}a}{2}q}\right]B^+ + e^{qd}\left[\gamma + \frac{2D}{J}\sin\left(\frac{k_x a}{2}\right)e^{\frac{\sqrt{3}a}{2}q}\right]B^- = 0 \tag{10b}$$

for zigzag edges, and

$$e^{q\left(d+\frac{a}{2\sqrt{3}}\right)}\cos\left(\frac{k_xa}{2}\right)A^+ + e^{-q\left(d+\frac{a}{2\sqrt{3}}\right)}\cos\left(\frac{k_xa}{2}\right)A^- + e^{qd}\left[\gamma + \frac{D}{J}\sin\left(\frac{k_xa}{2}\right)e^{q\frac{a\sqrt{3}}{2}}\right]B^+ +$$

$$e^{-qd}\left[\gamma + \frac{D}{J}\sin\left(\frac{k_xa}{2}\right)e^{-q\frac{a\sqrt{3}}{2}}\right]B^- = 0 \quad (11\text{a})$$

$$e^{-qd}\left[\gamma - \frac{D}{J}\sin\left(\frac{k_xa}{2}\right)e^{-\frac{\sqrt{3}a}{2}q}\right]A^+ + e^{qd}\left[\gamma - \frac{D}{J}\sin\left(\frac{k_xa}{2}\right)e^{\frac{\sqrt{3}a}{2}q}\right]A^+ + e^{-q\left(d+\frac{a}{2\sqrt{3}}\right)}\cos\left(\frac{k_xa}{2}\right)B^+ +$$

$$e^{q\left(d+\frac{a}{2\sqrt{3}}\right)}\cos\left(\frac{k_xa}{2}\right)B^- = 0 \quad (11\text{b})$$

for bearded edges. Additional details on the derivation of equations 10 and 11 are presented in Supplementary Note 2.

Equations 10 together with equations 6a and 6b (equivalently 6c and 6d) present a linear system of 4 equations in the coefficients $\{A^+, A^-, B^+, B^-\}$. For the system to admit non-zero solutions, the determinant of the defining matrix should be zero, which yields the final boundary condition equation for the ZE nanoribbon

$$2\gamma\left[-1 + \gamma\left(\frac{4D}{J}f_{DM} + 3\gamma - \Omega\right)\right]\cos\left(\frac{k_xa}{2}\right)\sinh\left[\frac{(\sqrt{3}a-12d)q}{6}\right]$$

$$+ \left[\frac{16D^2}{J^2}\gamma f_{DM}^2 + \Omega + \frac{4D}{J}f_{DM}(-1 + 5\gamma^2 - 2\gamma\Omega) + \gamma(6\gamma^2 - 5\gamma\Omega + \Omega^2) + 2\gamma\cos(k_xa) -\right.$$

$$\left.\frac{2D}{J}\sin(k_xa)\right]\sinh\left[\frac{(a+2\sqrt{3}d)q}{\sqrt{3}}\right]$$

$$+2\left\{\left(-\frac{4D}{J}f_{DM} - 2\gamma + \Omega\right)\cos\left(\frac{k_xa}{2}\right) + \frac{D}{J}\sin\left(\frac{k_xa}{2}\right)\left[-3 + \left(-\frac{4D}{J}f_{DM} - 3\gamma + \Omega\right)^2 - 2\cos(k_xa) + \right.\right.$$

$$\left.\left.\frac{2D}{J}\left(\frac{4D}{J}f_{DM} + 3\gamma - \Omega\right)\sin(k_xa)\right]\right\}\sinh\left[\frac{(5a+4\sqrt{3}d)q}{2\sqrt{3}}\right]$$

$$+\frac{2D}{J}\left[-\frac{D}{J}\left(\frac{4D}{J}f_{DM} + 3\gamma - \Omega\right)(-1 + \cos(k_xa)) - \sin(k_xa)\right]\sinh\left[\frac{(4a+2\sqrt{3}d)q}{\sqrt{3}}\right] = 0 \quad (12)$$

The boundary condition equation for BE nanoribbons, derived in the same way, reads

$$2\cos\left(\frac{k_xa}{2}\right)\left[\frac{16D^2}{J^2}\gamma f_{DM}^2 + \Omega + \frac{4D}{J}f_{DM}(-1 + 4\gamma^2 - 2\gamma\Omega) + \gamma(3\gamma^2 - 4\gamma\Omega + \Omega^2) + \right.$$

$$\left.\left(-\frac{4D}{J}f_{DM} - \gamma + \Omega\right)\cos(k_xa) - \frac{D}{J}\sin(k_xa)\right]\sinh\left[\frac{(\sqrt{3}a+12d)q}{6}\right]$$

$$+2\gamma\left[-1 + \gamma\left(\frac{4D}{J}f_{DM} + 3\gamma - \Omega\right) - \cos(k_xa)\right]\sinh\left[\frac{(a-2d)q}{\sqrt{3}}\right]$$

$$-\left\{\gamma-\Omega+\frac{D}{J}\left[-4\left(-1+\frac{D^2}{J^2}\right)f_{DM}+\frac{D}{J}(-3\gamma+\Omega)\right]+\left[\frac{4D}{J}f_{DM}+\frac{4D^3}{J^3}f_{DM}+\gamma+\frac{3D^2}{J^2}\gamma-\right.\right.$$

$$\left.\left(1+\frac{D^2}{J^2}\right)\Omega\right]\cos(k_xa)+\frac{D}{J}\left[3-\left(-\frac{4D}{J}f_{DM}-3\gamma+\Omega\right)^2+2\cos(k_xa)\right]\sin(k_xa)\right\}\sinh\left[\frac{(2a+2d)q}{\sqrt{3}}\right]$$

$$+\frac{2D}{J}\left[-\cos\left(\frac{k_xa}{2}\right)+\frac{D}{J}\left(\frac{4D}{J}f_{DM}+3\gamma-\Omega\right)\sin\left(\frac{k_xa}{2}\right)\right]\sin(k_xa)\sinh\left[\frac{(7a+4\sqrt{3}d)q}{2\sqrt{3}}\right]=0 \qquad (13)$$

This completes the classical field formalism for spin waves in the bounded nanoribbons. The real $q$ solutions of equations 12 and 13 determine the decay factors for the edge spin wave modes. Similarly, the imaginary solutions ($q=ik_y$) determine the allowed wavevectors for bulk modes. The energy dispersion curves for edge and bulk modes can then be determined using equation 8.

**Bulk and edge spin waves.** For numerical applications, we set $d$ as $\frac{35}{2\sqrt{3}}a$ and $\frac{37}{2\sqrt{3}}a$ for ZE and BE nanoribbons respectively. We note that the value of $d$ affects the number and characteristics of the bulk modes but has negligible effects on the edge modes.

To determine the allowed bulk spin waves, we set $q=ik_y$ in equations 12 and 13, where $k_y$ represents the wavevector component along the y-axis. Further, $\Omega$ is replaced by the solutions $\Omega_+$ or $\Omega_-$ from equation 8. The allowed $k_y$ are then obtained from the contour plots of equations 12 and 13. We find that $\Omega_+$ and $\Omega_-$ yield identical results for the allowed $k_y$ regardless of the DMI or magnetic anisotropy. An example of the $k_y$ solutions in the first BZ of the honeycomb lattice is presented in Fig. 3 (brown curves) for magnetically isotropic ($\gamma=1$) ZE nanoribbon with $D=0.1J$.

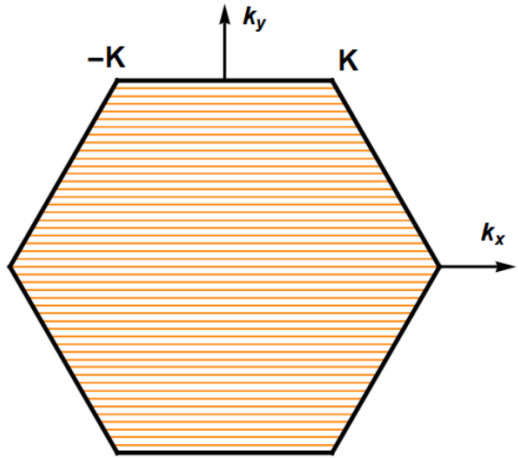

**Fig. 3:** The discretized $k_y$ solutions (brown curves) as functions of $k_x$ in the BZ of a magnetically ordered ZE nanoribbon with $D=0.1J$ and $d=35/2\sqrt{3}$. This nanoribbon hosts 49 bulk modes.

Like $k_y$, the decay factors $q$ for edge spin waves are determined from the contour plots of equations 12 and 13. Fig. 4 presents the edge spin waves decay factors for magnetically isotropic ZE and BE nanoribbons and selected values of the DMI coefficient. The decay factors $q$ are plotted as function of $k_x$ in the interval $|k_x| \leq \pi$. The green and dark cyan curves respectively represent the $q$ solutions corresponding to $\Omega_+$ and $\Omega_-$ energy solutions, which we will denote as $q_+$ and $q_-$ respectively.

Notably, the DMI is found to reduce the decay factors for edge spin waves near the BZ boundaries for both nanoribbon types. Additionally, the numerical plots show that BE nanoribbons present more exotic decay factors, notably the existence of $q$-solutions confined near the BZ boundaries. In the absence of the DMI, the $q_+$ and $q_-$ solutions are found to be degenerate (identical). Introducing the DMI lifts this degeneracy and elegantly combines the $q_+$ and $q_-$ solutions to form the continuous and differentiable (smooth) $q$-curves. As shown later, this DMI induced effect gives rise to the new class of nonreciprocal edge spin waves.

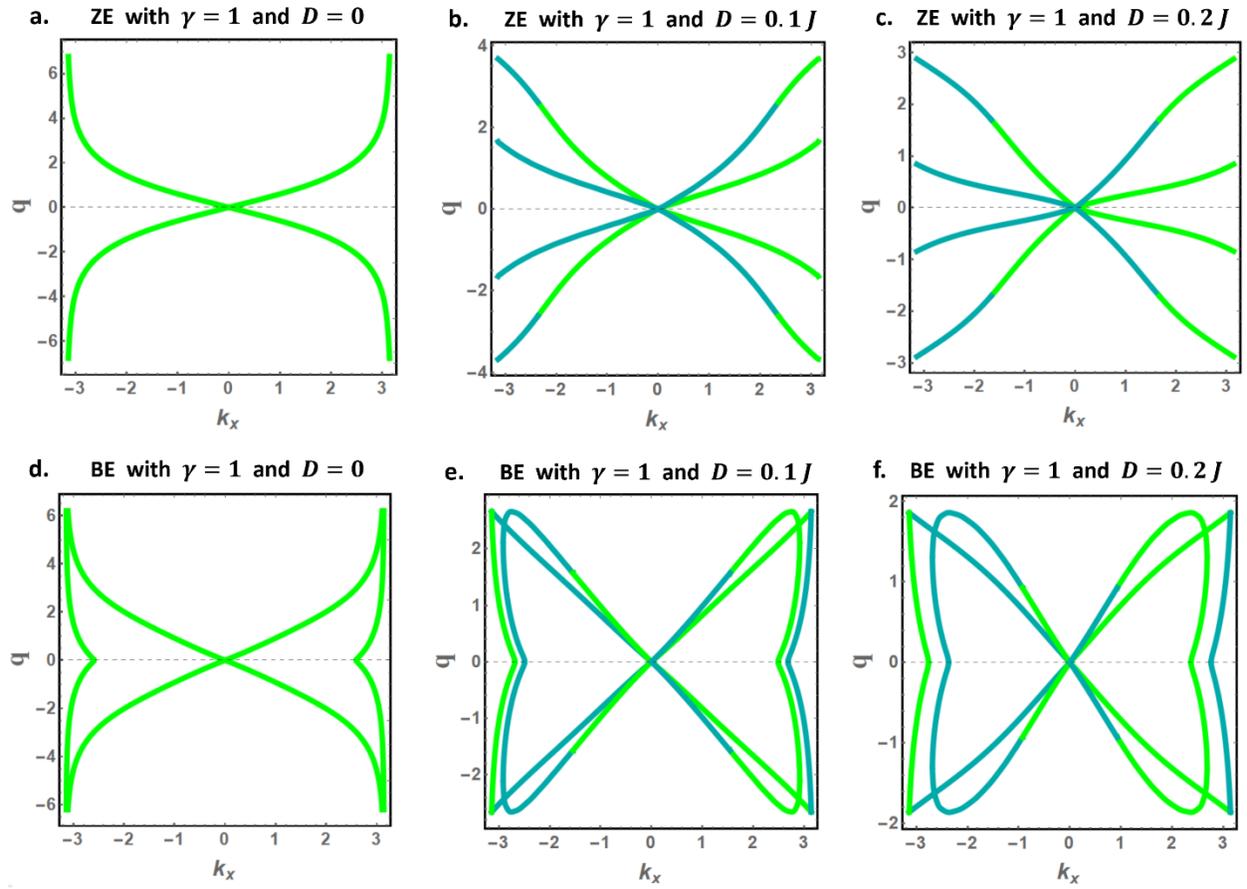

**Fig. 4:** Decay factors for edge spin waves in magnetically isotropic ZE and BE nanoribbons, (a, b, and c) for ZE, and (d, e, and f) for BE systems. The green and dark cyan curves represent the $q$ solutions corresponding to $\Omega_+$ and $\Omega_-$ energy solutions respectively.

With $k_y$ and $q$ solutions in hand, equation 8 can be used to determine the dispersion curves for bulk and edge spin waves. These are presented in Fig. 5. For bulk spin waves, we find that the $\Omega_+$ solution is positive (brown curves) whereas the $\Omega_-$ solution is negative for any allowed wavevector in the BZ, in agreement with the conventional results on bulk spin waves in an antiferromagnet [11, 51, 52, 58, 59, 61]. The $\Omega_-$ solution is hence excluded for bulk spin waves, leaving the $\Omega_+$ solution as the unique physical solution. The numerical results further show the evolution of the discretized bulk spin waves band as the strength of the DMI increases. The DMI induces nonreciprocal bulk spin waves with a unique polarization (conventional nonreciprocal spin waves).

Before turning to the edge spin waves dispersion, we analyze the topology of the discretized propagation band. The Chern number can be calculated using the equation

$$C = \eta \iint dk_x dk_y Tr\left[P\left(\frac{\partial P}{\partial k_x}\frac{\partial P}{\partial k_y} - \frac{\partial P}{\partial k_y}\frac{\partial P}{\partial k_x}\right)\right]$$

The term inside the trace represents the Berry curvature, $\eta$ is a normalization constant and the integral is over the allowed $k_x$ and $k_y$ in the first BZ. The matrix $P = |V_+(\vec{k})\rangle\langle V_+(\vec{k})|$ is the projection matrix and $|V_+(\vec{k})\rangle$ represents the normalized eigenvector of the matrix formed by equations 6a and 6c. This eigenvector particularly corresponds to the eigenvalue $\Omega_+$. Finally, $\langle V_+(\vec{k})|$ is the Hermitian conjugate of $|V_+(\vec{k})\rangle$.

In the Néel order phase, the DMI does not contribute to the Berry curvature. The Chern number calculation for nanoribbons with or without DMI yields 0,

$$C = \eta \iint dk_x dk_y \frac{i\left[-\cos\left(\frac{k_x}{2}\right) + \cos\left(\frac{\sqrt{3}k_y}{2}\right)\right]\sin\left(\frac{k_x}{2}\right)}{6\sqrt{3}\gamma\sqrt{-3 + 9\gamma^2 - 2\cos(k_x) - 4\cos\left(\frac{k_x}{2}\right)\cos\left(\frac{\sqrt{3}k_y}{2}\right)}} = 0$$

and the discretized propagation band is hence topologically trivial.

We now consider edge modes. For the magnetically isotropic case, both nanoribbon types host edge spin waves with linear dispersion curves near the BZ center, even in the presence of the DMI. This generalizes the predictions of previous theoretical studies on spin waves in bounded 2D honeycomb antiferromagnets with zero DMI [26, 39, 52]. In this context, we note that the results for isotropic ZE nanoribbon with zero DMI (Fig. 5a) are consistent with a previous study addressing spin waves in semi-infinite 2D honeycomb antiferromagnets [26].

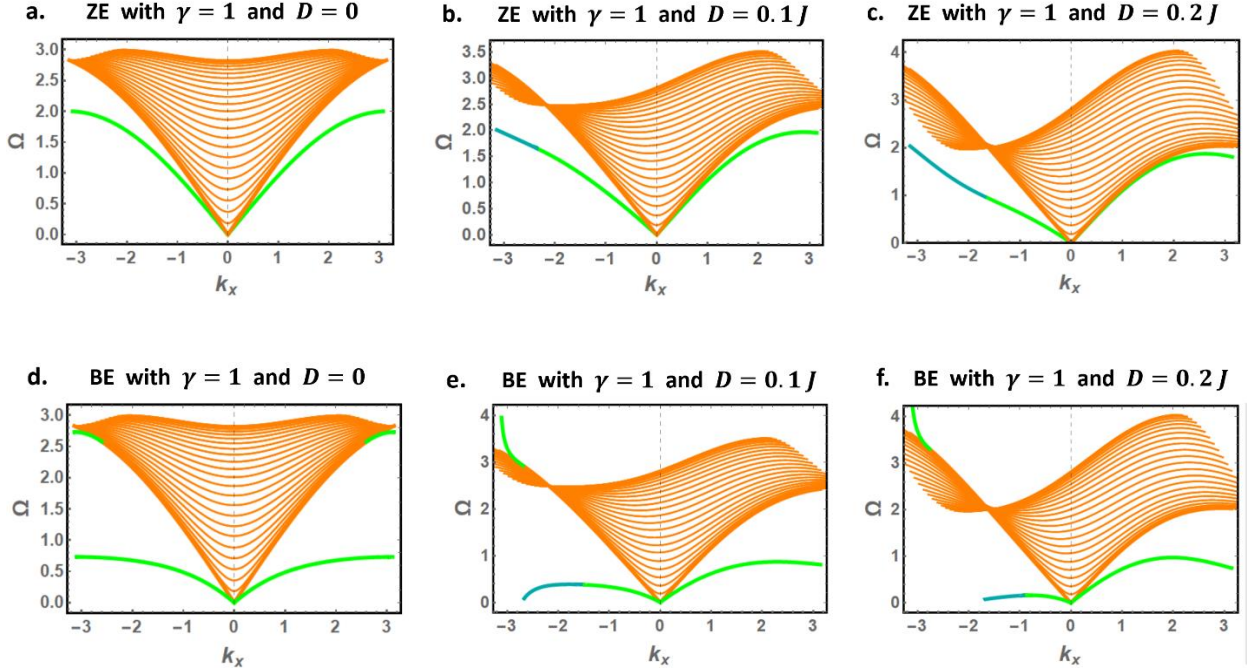

**Fig. 5:** The physically accepted dispersion curves for bulk (brown curves) and edge (green and dark cyan curves) spin waves in magnetically isotropic nanoribbons with zigzag (a, b, and c) and bearded (d, e, and f) edge boundaries.

Unlike bulk spin waves, both $\Omega_+$ and $\Omega_-$ conspire to give the physically accepted dispersion curves for edge spin waves in nanoribbons with DMI. These solutions are found to switch from $\Omega_+$ to $\Omega_-$ at a critical $k_x$ value, denoted $k_{x,c}$. The value of $k_{x,c}$ shifts toward the origin ($k_x = 0$) as the strength of the DMI increases, allowing a broader contribution of the $\Omega_-$ solution. Moreover, the DMI positive contribution to $\Omega_-$ is found to dominate the exchange negative contribution in ZE nanoribbons. Consequently, $\Omega_-$ remains positive up to the BZ boundary in ZE nanoribbons (see equation 8). For BE nanoribbons, however, the $\Omega_-$ dispersion curve is found to bend towards negative values at large $|k_x|$.

The eigenvectors of the matrix defined by equations 10a, 10b, 6a and 6b demonstrate another interesting phenomenon, namely that the edge spin waves propagate in opposite directions on the opposite edges of the ZE nanoribbons (similarly, for the BE nanoribbons). This agrees with our recent study [52], limited to large wave length spin waves without the DMI. The symmetry behind this phenomenon is discussed extensively in [52]. Combining all these results, we deduce that for $|k_x| \geq -k_{x,c}$, spin waves on the two edges of the nanoribbon propagate in opposite directions, with opposite polarization and different energies. A graphical illustration of this new class of nonreciprocal edge spin waves is presented in Fig. 6.

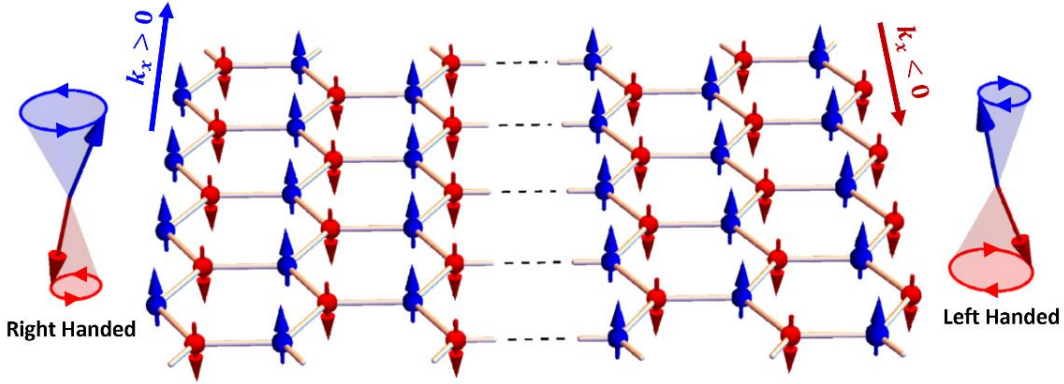

**Fig. 6:** Schematic illustration of the DMI induced new class of nonreciprocal spin waves propagating in opposite directions and polarizations at the edges of a ZE nanoribbon.

We proceed and discuss another striking effect induced by the DMI on the high-energy edge spin waves, present only in BE nanoribbons. These dispersion curves correspond to the $q$-solutions confined near the BZ boundaries in Figs. 4d, 4e, and 4f. For DMI free nanoribbons (Fig. 5d), the high-energy edge dispersion curves are located just below the propagating band and on both BZ boundaries. In the presence of the DMI (Figs. 5e and 5f), the right dispersion curve is no more physical (shifts to negative energies) while the left dispersion curve is lifted above the propagation band by the DMI. The DMI further alters the characteristics of the left dispersion curve, now characterized by exceptionally large group velocities near the BZ boundary. Recalling that the BE nanoribbon allows only counter-propagating spin waves on opposite edges, these unidirectional edge spin waves propagate exclusively on one edge of the nanoribbon.

We next analyze the effects of magnetic anisotropy on the edge and bulk spin waves. As in Figs. 4 and 5, we present in Figs. 7 and 8 the decay factors and the physically accepted dispersion curves for bulk and edge spin waves in magnetically anisotropic nanoribbons ($\gamma = 1.1$). Remarkably, all previously stated conclusions survive the effect of anisotropy. Furthermore, magnetic anisotropy is found to induce important energy gaps between the low-energy edge spin waves and the propagating spin waves throughout the BZ.

Exceptionally interesting are the low-energy edge spin waves in BE nanoribbons (Figs. 8d, 8e and 8f) which can be excited separate from the bulk spin waves throughout the BZ. Moreover, the numerical results in the presence of the DMI (Figs. 8e and 8f) demonstrate the possibility of controlling the propagation direction and the polarization of these edge modes. To illustrate, we highlight that edge spin waves with $k_x > 0$ has higher energies compared to their counterparts with $k_x < 0$. Consequently, with appropriately tuned energy $\Omega > \Omega_0 = \Omega(k_x = 0)$, the nanoribbon allows unidirectional edge mode with $k_x > 0$ on one of its edges. Tuning the energy below $\Omega_0$ flips the direction of the edge spin wave while preserving or reversing the polarization, depending whether $k_x > k_{x,c}$ or $k_x < k_{x,c}$ respectively. These nanoribbons are hence predicted to be model candidates for the realization of magnetic insulators with technologically desired features.

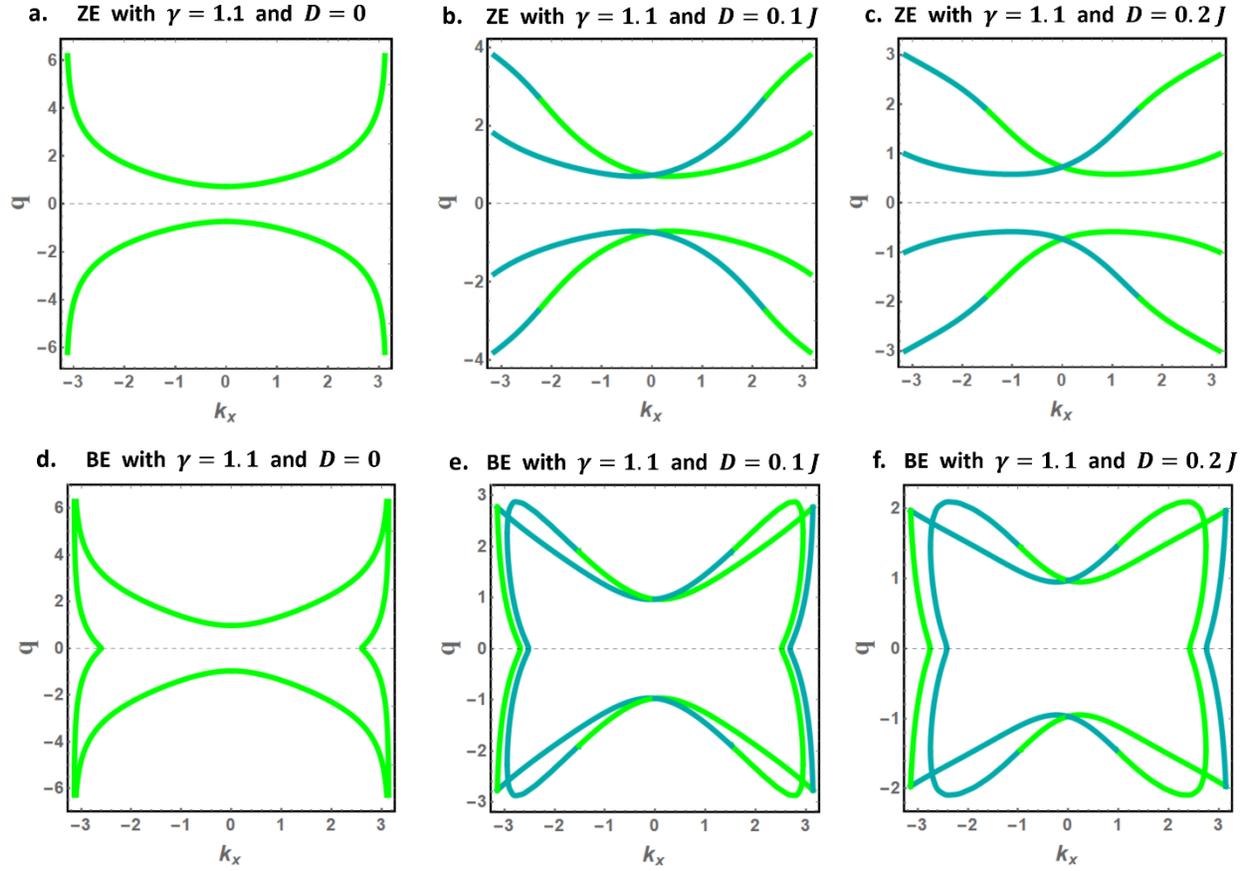

**Fig. 7:** Decay factors for edge spin waves in magnetically anisotropic nanoribbons ($\gamma = 1.1$) with zigzag (a, b, and c) and bearded (d, e, and f) edge boundaries.

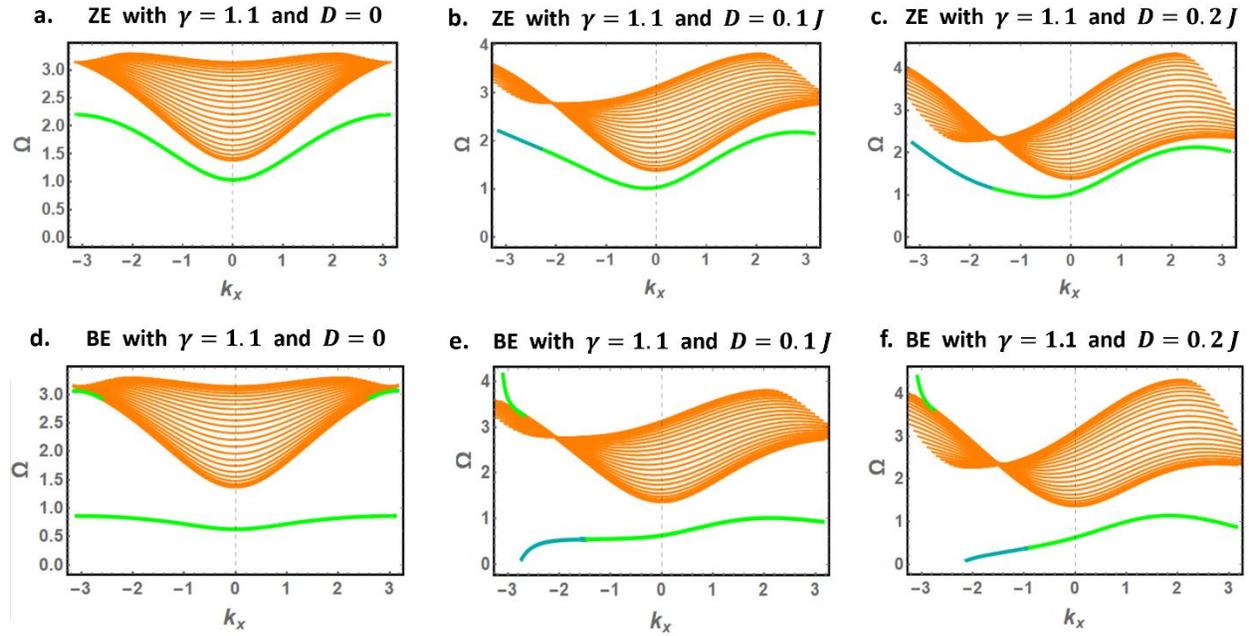

**Fig. 8:** The physically accepted dispersion curves for bulk and edge spin waves in magnetically anisotropic ($\gamma = 1.1$) nanoribbons with zigzag (a, b, and c) and bearded (d, e, and f) edge boundaries.

We close this section with a brief discussion on the dispersion curves formed as a combination of $\Omega_+$ and $\Omega_-$ solutions, leading to unconventional nonreciprocal spin waves. We first recall that dispersion curves in a wave theory are generally required to be continuous and differentiable (smooth). This ensures that the excitation's energy does not change abruptly after an infinitesimal change in the wavevector. The solution $\Omega_+$ satisfies these conditions as long as the square root function $\sqrt{9\gamma^2 - f_{ex}^+ f_{ex}^-}$ does not vanish in the BZ, which is the case for the bulk modes. Remarkably, for edge spin waves with DMI, $k_{x,c}$ is a root of $\sqrt{9\gamma^2 - f_{ex}^+ f_{ex}^-}$. Consequently, the continuous and differentiable dispersion function can only be formed as a combination of $\Omega_+$ and $\Omega_-$ solutions.

## Discussions

We have investigated the exotic physics and technological potentials of magnetic excitations in 2D honeycomb antiferromagnetic nanoribbons. We predict a new class of nonreciprocal edge spin waves characterized by opposite polarizations when propagating in opposite directions. These remarkable spin waves are induced by the DMI and are found to satisfy key requirements for applications in antiferromagnetic magnonics.

In particular, our results constitute an important contribution to the current efforts seeking to establish unconventional magnonic devices utilizing spin wave polarization in antiferromagnets. Unlike ferromagnets, collinear antiferromagnets host two spin wave modes with opposite polarization and provide the opportunity to encode magnonic information in the polarization degree of freedom. The realization of polarization-based magnonic circuits, however, requires an efficient way to manipulate the polarization and the transmission direction of spin waves. In view of our results, 2D honeycomb antiferromagnetic nanoribbons are predicted to serve this purpose.

Particularly interesting are the anisotropic BE nanoribbons, presenting low-energy edge spin waves that can be excited throughout the BZ, at energies below the propagation band. The edge sites of BE nanoribbons can hence be used as natural waveguide to transport spin waves along well-defined and narrow paths. For zero DMI, these spin waves propagate in opposite directions on the opposite edges of the BE nanoribbons. The DMI confines these modes to one of the edges and offers the practical opportunity to excite unidirectional spin waves with a pre-selected polarization. BE nanoribbons with DMI also host unidirectional high-frequency spin waves, propagating with large group velocities on one of the nanoribbons edges. These are indeed unconventional edge excitations which are also potentially interesting for magnonic information processing.

Among the various types of boundary magnetic excitations (surface, interface and domain wall spin waves), edge spin waves in 2D magnetic materials are apparently the most exotic. The predictions of the current studies can indeed be tested in realistic 2D honeycomb antiferromagnets, such as Mn-based trichalcogenide [36]. This definitely requires experimental efforts to fabricate monolayer nanoribbons with appropriate edge structure and magnetism.